# Structural and Electronic Properties of Graphene and Graphene-like Materials

Gautam Mukhopadhyay[*] and Harihar Behera

*Department of Physics, Indian Institute of Technology Bombay, Powai, Mumbai-400076, India*

[*]*Corresponding author, E-mail: gmukh@phy.iitb.ac.in*

**Abstract:** Using full potential density functional theory calculations we have investigated the structural and electronic properties of graphene and some other graphene-like materials, viz., monolayer of SiC, GeC, BN, AlN, GaN, ZnO, ZnS and ZnSe. We hope, with the advancement of material synthesis techniques, some these new materials will be synthesized in the near future for potential applications in various nano-devices.

## Introduction

Graphene [1], a one-atom-thick crystal of $sp^2$-bonded carbon atoms in a two dimensional (2D) honeycomb lattice is nothing but a monolayer of graphite (ML-C). The exotic mechanical, thermal, electronic, optical and chemical properties of graphene promise many novel applications [1]. Recently, the phenomenal growth in the research on graphene has inspired the study of other graphene-like 2D materials [1-6]. A number of 2D/quasi-2D nanocrystals of BN [1], $MoS_2$ [1], $MoSe_2$ [1], $Bi_2Te_3$ [1], Si [2], ZnO [3] have been synthesized. Using density functional theory (DFT) calculations, recently Freeman et al [4] predicted that when the layer number of (0001)-oriented wurtzite (WZ) materials (e.g., AlN, BeO, GaN, SiC, ZnO and ZnS) is small, the WZ structures transform into a new form of stable hexagonal BN-like structure. This prediction has recently been confirmed for ZnO [3], whose stability is attributed to the strong in-plane $sp^2$ hybridized bonds between Zn and O atoms. Graphene-like 2D/quasi-2D honeycomb structures of group-IV and III-V binary compounds have also been studied [5, 6] by DFT calculations. Here, we briefly present our calculations on the structural and electronic properties of graphene and some other graphene-like monolayer (ML) structures of the binary compounds, viz., BN, AlN, GaN, ZnO, ZnS, SiC, GeC, using the DFT based full-potential (linearized) augmented plane wave plus local orbital (FP-(L)APW+lo) method [7].

## Calculation methods

We use the elk-code [8] and the Perdew-Zunger variant of LDA [9] for our calculations, the accuracy of which has been successfully tested in our previous studies [10-12]. For plane wave expansion in the interstitial region, we have used $8 \leq |\mathbf{G}+\mathbf{k}|_{max} \times R_{mt} \leq 9$ where $R_{mt}$ is the smallest muffin-tin radius, for deciding the plane wave cut-off. The Monkhorst-Pack [13] **k**-point grid size of 20×20×1 was used for structural and of 30×30×1 for band structure calculations. The total energy was converged within 2μeV/atom. We simulate the 2D-hexagonal structure as a 3D-hexagonal supercell with a large value of *c*-parameter (= |**c**| = 40 a.u.). Inset of Figure 1 depicts the top-down view of a ML-BN in planar graphene-like honeycomb structure which is the prototype of all the materials considered here (for graphene, B and N atoms are to be replaced by the C atoms).

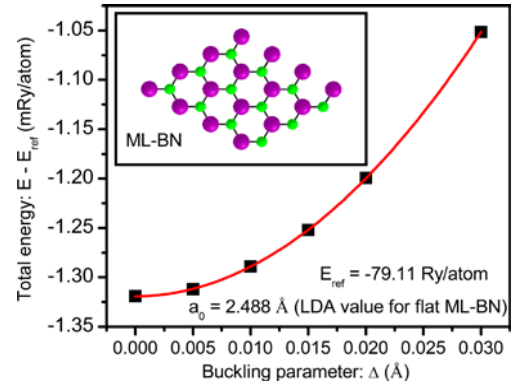

**Figure 1.** Probing buckling in ML-BN using the principle of minimum energy for the stable structure; in buckled ML-BN, B and N atoms take positions in two different parallel planes, buckling parameter 'Δ' is the perpendicular distance between those parallel planes and Δ = 0.0 Å for a flat ML-BN. Inset shows the top-down view of a flat ML-BN in ball-stick model.

## Results and discussions

For an assumed flat ML-BN, our calculated LDA value of ground state in-plane lattice constant $a_0$ = 2.488 Å agrees with the reported values [5, 6]. Our assumption on the flat ML-BN structure was tested as correct by the calculated results depicted Figure 1. The principle used in Figure 1 was also applied to all the materials listed in Table 1 (except ML-ZnO and ML-ZnSe which are being probed now) to investigate if these structures adopt buckled structure like silicene (graphene-analogue of Si) [2, 14]. Our calculated structural parameters and the band gaps of ML-C, ML-SiC, ML-GeC, ML-BN, ML-AlN, ML-GaN, ML-ZnO, ML-ZnS and ML-ZnSe are all listed in Table 1, which are in agreement with the available theoretical data [5, 6]. In Figure 2 we present the band structures of these materials as calculated within LDA. We note that except for ML-C, actual band gaps listed in Table 1 would be more as LDA underestimates the gap.

**Table 1.** Calculated LDA values of ground state in-plane lattice constants $a_0$ (=|**a**|=|**b**|), buckling parameters $\Delta$ and band gaps $E_g$ of monolayer graphene (ML-C) and some structurally similar binary compounds.

| Material | $a_0$ (Å) | $\Delta$ (Å) | $E_g$ (eV) |
|---|---|---|---|
| ML-C | 2.445 | 0.0 | 0.000 (K→K) |
| ML-SiC | 3.066 | 0.0 | 2.547 (K→M) |
| ML-GeC | 3.195 | 0.0 | 2.108 (K→K) |
| ML-BN | 2.488 | 0.0 | 4.606 (K→K) |
| ML-AlN | 3.090 | 0.0 | 3.037 (K→Γ) |
| ML-GaN | 3.156 | 0.0 | 2.462 (K→Γ) |
| ML-ZnO | 3.20 | 0.0 (used) | 1.680 (Γ→Γ) |
| ML-ZnS | 3.7995 | 0.0 | 2.622 (Γ→Γ) |
| ML-ZnSe | 3.996 | 0.0 (used) | 1.866 (Γ→Γ) |

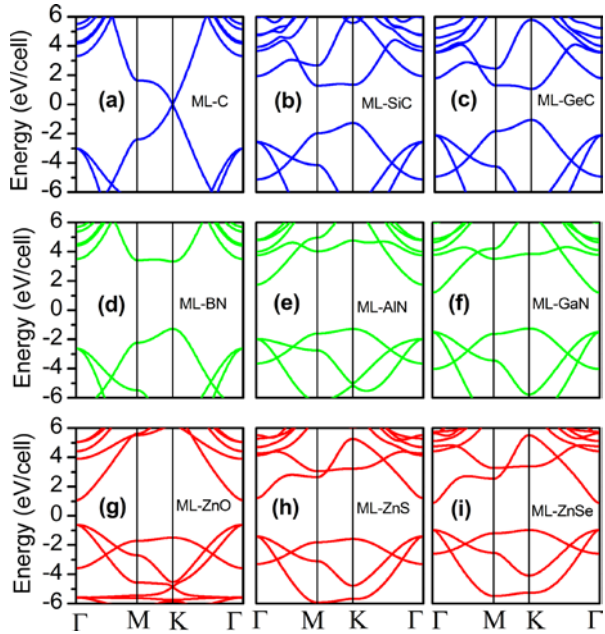

**Figure 2.** Band structures (LDA) of graphene (ML-C) and mono-layers of SiC, GeC, BN, AlN, GaN, ZnO, ZnS and ZnSe in graphene-like planar structure

## Conclusions

Using full potential DFT calculations we have investigated the structural and electronic properties of graphene and some other graphene-like materials. While our results corroborate the previous theoretical studies based on different methods, our calculations on ML-ZnSe is something new. We hope, with the advancement of fabrication techniques, the hypothetical graphene-like materials discussed here will be synthesized in the near future for potential applications in a variety of novel nano-devices.

## References

1. Neto A. H. C., and Novoselov K. New directions in science and technology: two-dimensional crystals, Rep. Prog. Phys. **74** (2011) 1-9.
2. Kara A., Enriquez H., et al. A review on silicene – New candidate for electronics, Surf. Sci. Rep. **67** (2012) 1-18.
3. Tusche C., et al. Observation of depolarized ZnO(0001) monolayers: Formation of unreconstructed planar sheets, Phys. Rev. Lett. **99** (2007) 026102.
4. Freeman C.L., Claeyssens F., et al. Graphitic nanofilms as precursors to wurtzite films: Theory, Phys. Rev. Lett. **96** (2006) 066102.
5. Şahin H., et al. Monolayer honeycomb structures of group-IV and III-V binary compounds: First-principles calculations, Phys. Rev. B **80** (2009) 155453.
6. Wang S. Studies of Physical Properties of Two-Dimensional Hexagonal Crystals by First-Principles Calculations, J. Phys. Soc. Jpn. **79** (2010) 064602.
7. Sjöstedt E., et al. An alternative way of linearizing the augmented plane-wave method, Solid State Commun. **114** (2000) 15-20.
8. Elk open source code: http://elk.sourceforge.net/
9. Perdew P., Zunger A. Self-interation correction to density-functional approximations for many electron systems, Phys. Rev. B **23** (1981) 5048.
10. Behera H., Mukhopadhyay G. Structural and electronic properties of graphene and silicene, AIP Conf. Proc.**1313** (2010) 152-155; arXiv:1111.1282
11. Behera H., Mukhopadhyay G. First-Principles Study of Structural and Electronic Properties of Germanene, AIP Conf. Proc.**1349** (2011) 823-824; arXiv:1111.6333
12. Behera H., Mukhopadhyay G. Strain-tunable band gap in graphene/h-BN hetero-bilayer, J. Phys. Chem. Solids **73** (2012) 818-821; arXiv:1204.2030
13. Behera H., Mukhopadhyay G. Strain-tunable band gaps of Two dimensional Hexagonal BN and AlN: An FP-(L) LAPW+lo study, AIP Conf. Proc. **1447** (2012) 273-274; arXiv:1206.3162v1
14. Monkhorst H.J., Pack J.D. Special points for Brillouin-zone integrations, Phys. Rev. B **13** (1976) 5188-5192.
15. Behera H., Mukhopadhyay G. A comparative computational study of the electronic properties of planar and buckled silicene, arXiv:1201.1164v1